\documentclass[aps,prl,twocolumn,superscriptaddress,preprintnumbers,floatfix,10pt,nofootinbib]{revtex4-2}

\usepackage[bookmarks=false]{hyperref}
\usepackage{amsmath}
\usepackage{graphicx}
\usepackage{xspace}
\usepackage[usenames,dvipsnames]{xcolor}
\usepackage[export]{adjustbox}
\usepackage{bm}
\usepackage{graphicx}
\usepackage{amsmath,amssymb,mathtools} 
\usepackage{xspace}
\usepackage{wrapfig}
\usepackage{slashed}
\usepackage[normalem]{ulem}
\usepackage[dvipsnames]{xcolor}
\usepackage[utf8]{inputenc}
\usepackage{mathtools}
\usepackage{stmaryrd}
\usepackage{appendix}
\usepackage{xspace}
\usepackage{upgreek}
\usepackage{multirow}

\usepackage{scrextend}

\usepackage{tabularx}
\usepackage{tensor}
\usepackage{soul}

\newcommand{\mb}[1]{\boldsymbol{#1}}

\definecolor{darkgreen}{rgb}{0.0, 0.4, 0.0}

\usepackage{marginnote}
\usepackage[normalem]{ulem}

\newcommand{\abs}[1]{\lvert#1\rvert}

\newcommand{\df}{\mathrm{d}}

\newcommand{\bn}{{\bar{n}}}

\newcommand{\ra}{\rightarrow}

\newcommand{\as}{\alpha_s}

\newcommand{\eps}{\epsilon}

\newcommand{\cB}{{\mathcal B}}

\newcommand{\cI}{{\mathcal I}}
\newcommand{\cN}{{\mathcal N}}

\newcommand{\cP}{{\mathcal P}}

\newcommand{\spl}{{\mathbf{Sp}}}

\newcommand{\nslash}{n\!\!\!\slash}
\newcommand{\bnslash}{\bar{n}\!\!\!\slash}

\def\dfbar{{\:\mathchar'26\mkern-12mu \df}}

\newcommand{\nn}{\nonumber}

\newcommand{\br}[1]{\left\langle #1 \left |}
\newcommand{\kt}[1]{\right | #1\right \rangle }
\newcommand{\brac}[1]{\left (#1\right )}

\newcommand{\letter}{\emph{Letter}\xspace}

\def\ln{\textrm{ln}}
\def\df{\textrm{d}}
\def\nn{\nonumber}

\allowdisplaybreaks[2]

\newcommand{\im}{\mathrm{i}}

\newcommand{\cO}{\mathcal{O}}

\usepackage[mathscr]{eucal}

\newcommand{\cM}{\mathcal{M}}

\DeclareRobustCommand{\Refcite}[1]{Ref.~\cite{#1}}

\DeclareRobustCommand{\Refscite}[1]{Refs.~\cite{#1}}

\DeclareRobustCommand{\eq}[1]{eq.~\eqref{eq:#1}}

\DeclareRobustCommand{\eqs}[2]{eqs.~\eqref{eq:#1} and \eqref{eq:#2}}

\DeclareRobustCommand{\secn}[1]{\hyperref[sec:#1]{section~\ref*{sec:#1}}}
\DeclareRobustCommand{\Sec}[1]{\hyperref[sec:#1]{Section~\ref*{sec:#1}}}

\DeclareRobustCommand{\subsec}[1]{\hyperref[subsec:#1]{subsection~\ref*{subsec:#1}}}
\DeclareRobustCommand{\Subsec}[1]{\hyperref[subsec:#1]{Subsection~\ref*{subsec:#1}}}

\DeclareRobustCommand{\app}[1]{\hyperref[app:#1]{appendix~\ref*{app:#1}}}
\DeclareRobustCommand{\App}[1]{\hyperref[app:#1]{Appendix~\ref*{app:#1}}}
\DeclareRobustCommand{\apps}[2]{appendices~\ref{app:#1} and \ref{app:#2}}

\DeclareRobustCommand{\fig}[1]{\hyperref[fig:#1]{figure~\ref*{fig:#1}}}
\DeclareRobustCommand{\Fig}[1]{\hyperref[fig:#1]{Figure~\ref*{fig:#1}}}

\DeclareRobustCommand{\tab}[1]{\hyperref[tab:#1]{table~\ref*{tab:#1}}}
\DeclareRobustCommand{\Tab}[1]{\hyperref[tab:#1]{Table~\ref*{tab:#1}}}

\usepackage[textsize=tiny,textwidth=38pt]{todonotes}

\allowdisplaybreaks[2]

\providecommand{\sectionPaper}[1]{\textit{\textbf{#1.}}}
\providecommand{\headingAcknowledgments}{\textit{\textbf{Acknowledgments. }}}
\providecommand{\citeSupplementBibEntry}{\cite{*[{See Supplemental Material at \textcolor{red}{\texttt{<insert url>}}}] [{}] supplement}}
\providecommand{\app}[1]{\citeSupplementBibEntry}
\providecommand{\apps}[2]{\citeSupplementBibEntry}

\begin{document}

\preprint{\vbox{\hbox{DESY-26-091}}}
\preprint{\vbox{\hbox{SLAC-PUB-260713}}}

\title{Collinear Factorization Violation and Reggeization}

\author{Damiano Barcaro}
\email{damiano.barcaro@desy.de}
\affiliation{Deutsches Elektronen-Synchrotron DESY, Notkestr. 85, 22607 Hamburg, Germany}

\author{Anjie Gao}
\email{anjiegao@stanford.edu}
\affiliation{SLAC National Accelerator Laboratory, Stanford University, Stanford, CA 94039, USA}

\author{Jonathan R. Gaunt}
\email{gaunt.jonathan@ucy.ac.cy}
\affiliation{Department of Physics, University of Cyprus, Nicosia 1678, Cyprus}

\author{Aditya Pathak}
\email{aditya.pathak@desy.de}
\affiliation{Deutsches Elektronen-Synchrotron DESY, Notkestr. 85, 22607 Hamburg, Germany}

\date{\today}

\begin{abstract}

We derive an all-order soft-collinear subgraph factorization in the space-like collinear limits of amplitudes for single gluon pinched in the Glauber region. We show that these contributions exhibit Reggeization, combining aspects of both forward and hard scattering.
By deforming away from the Glauber region wherever possible we demonstrate a dramatic simplification of the effective theory subgraphs and
thereby drastically reducing the complexity in computing collinear factorization breaking terms in multi-point amplitudes.
Our work is aided with development of new software tools initiating a systematic study of processes involving Glauber exchanges at a multi-loop level.
\end{abstract}

\maketitle

\sectionPaper{Introduction}
Renormalizable gauge theories display a rich structure of the infrared singularities and associated logarithms. Establishing soft-collinear factorization of gauge theory amplitudes and cross sections requires demonstrating the non-existence or eventual cancellation of contributions from the so-called `Glauber regions'. This has been shown to be the case for sufficiently inclusive processes in hadron colliders~\cite{Bodwin:1984hc,Collins:1985ue,Collins:1988ig, Collins:2011zzd, Diehl:2015bca}. Beyond this, factorization violation has been established in several cases~\cite{Rogers:2010dm,Catani:2011st,Forshaw:2012bi,Gaunt:2014ska,Zeng:2015iba, Schwartz:2017nmr,Schwartz:2018obd,Cieri:2024ytf,Henn:2024qjq,Guan:2024hlf,Duhr:2025lyg, Dasgupta:2025cgl, Banfi:2025mra, Becher:2026kbr}.
Given the current precision physics programme at the LHC, crucial questions of significant current interest are where and how factorization is generically violated, and if the lowest-scale physics nonetheless still factorizes into the PDFs in these cases~\cite{Becher:2024kmk,Becher:2025igg,Becher:2026kbr,Banfi:2025mra}.

Such questions cannot be fully addressed through analyses at the lowest loop order where the Glauber region is first revealed. Instead, a systematic expedition into the multi-loop territory is required, as each loop order can potentially reveal new features. This is best seen in the ``space-like collinear limits'' of amplitudes when an outgoing parton is made collinear to an incoming parton. Here, the same Glauber mechanism leads to the (strict) collinear factorization violation (CFV) where the splitting amplitudes develop dependence on the color charges and momenta of non-collinear partons~\cite{Catani:2011st,Forshaw:2012bi}. At one-loop, this is a simple phase (see \eq{sp1cfv}), but at two-loops \Refcite{Catani:2011st} showed using the known structure of infrared (IR) poles of amplitudes~\cite{Sotiropoulos:1993rd,
Korchemsky:1993hr,
Korchemskaya:1994qp,
Korchemskaya:1996je,
Catani:1998bh,
Aybat:2006wq,
Aybat:2006mz,
Dixon:2008gr,
Becher:2009cu,
Gardi:2009qi,
Becher:2009qa,
Gardi:2009zv,
Dixon:2009ur,
DelDuca:2011xm,
DelDuca:2011ae,
Caron-Huot:2013fea,
DelDuca:2013ara,
DelDuca:2014cya,
Almelid:2015jia,
Caron-Huot:2017fxr,
Almelid:2017qju,
Caron-Huot:2017zfo}
that a new logarithm of cross ratio of kinematic invariants emerges (see \eq{sp2cfv}).
However, while the IR factorization formula can predict the poles in the CFV pieces, it offers little explanation from the perspective of the IR regions exhibiting the Glauber mechanism.
Such an understanding is imperative to investigate, for example, CFV at hadron colliders arising from spectator-spectator interactions~\cite{Duhr:2025lyg}.
Finally, such CFV pieces cannot be inferred from time-like kinematics and must be extracted from considering loop corrections to the full multi-point amplitude, making their explorations directly from the fixed order calculations challenging.
The two-loop piece was computed for $\cN = 4$ supersymmetric Yang-Mills in \Refcite{Henn:2024qjq} and for QCD only very recently in \Refcite{Buccioni:2026mfg}.

\begin{figure}[t]
\centering
\includegraphics[width=0.3\textwidth]{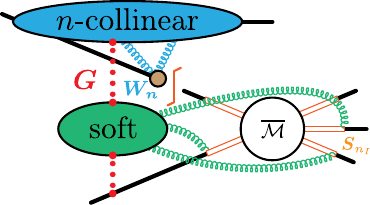}
\caption{A generic collinear and soft subgraph decomposition of collinear factorization violating pieces for a single Glauber exchange (red-dotted). The brown blob represents emissions that can be absorbed into a collinear Wilson line.}
\label{fig:fact}
\end{figure}

In this \letter we approach this problem from the perspective of the effective field theory (EFT) that captures the IR regions underpinning the CFV.
Concretely, we show that the two-loop CFV piece exhibits \textit{a large rapidity logarithm} which
finds its origin in an interesting mix of the Reggeization phenomenon associated with forward scattering and contributions from collinear and soft Wilson lines in the hard scattering.
We introduce a new approach for isolating Glauber regions via SCET Glauber operators~\cite{Rothstein:2016bsq} that preserves analyticity. This enables us to expose remarkable simplifications in the organization of the EFT subgraphs
leading to the factorized picture shown in \fig{fact} for a single gluon pinched in the Glauber region.
Our work lays the foundation for
understanding the hitherto-unknown ``bottom-up'' factorization and resummation of observables such as gap-between-jets that expose the Glauber phases and Glauber regions as super-leading logarithms~\cite{Becher:2021zkk,Becher:2023mtx}.

The outline of this \letter is as follows: after a brief review of the known CFV pieces and the formalism of Glauber SCET, we describe the factorization of subgraphs depicted in \fig{fact}. The one-loop collinear subgraphs will be shown to exhibit rapidity divergences combining to give BFKL-ladder like color structures needed to cancel the rapidity poles in the soft subgraph. Finally, we will connect these findings with the exponentiation of the rapidity logarithm using the IR factorization formula. We leave the technical details and the computation of the two-loop finite CFV pieces to a companion paper~\cite{cfv_long}.

\sectionPaper{CFV in amplitudes}
An amplitude $p_1p_3\! \ra\! p_2 p_4 p_5\!\ldots$ in the space-like collinear limit $p_1 || p_2$ factorizes as
\begin{align}
\big | \cM(p_1,p_2,\ldots) \big] &\overset{p_1 || p_2}{\simeq} \Big(\spl^{(0)}(p_1,p_2; \tilde P) + \frac{\as}{2\pi}\spl^{(1)} \\
& \nn+\brac{\frac{\as}{2\pi}}^2 \spl^{(2)} + \ldots\Big)\big | \cM(\tilde P, p_3, \ldots) \big] \,,
\end{align}
where $\spl$ are color operators acting on $\big|\cM\big]$ with unspecified color. Here $\tilde P \simeq p_1 + p_2$ and $\tilde P^2 = 0$.
\Refscite{Catani:2011st,Forshaw:2012bi} showed that splitting amplitudes $\spl^{(i)}$ develop dependence at loop-levels on color charges and momenta of other non-collinear partons.
At one-loop, one finds a new correction that appears only for space-like splitting:
\begin{align}\label{eq:sp1cfv}
\spl^{(1)}_\text{cfv}
=\mb \Delta^{(1)}_C\spl^{(0)} = -\frac{2 \pi\im}{\eps}\,\mb T_2 \cdot\mb T_{\rm in} \spl^{(0)} + \cO(\eps)\, ,
\end{align}
where $\mb T_{\rm in} \equiv \mb T_3$.
At two-loops, exploiting the known properties of infrared (IR) poles of amplitudes, one finds
\begin{align}\label{eq:sp2cfv}
\spl^{(2)}_\text{cfv} =\mb \Delta^{(2)}_C \spl^{(0)} + \ldots , \quad\mb \Delta^{(2)}_C = \frac{1}{2} \left[\mb I_{\overline \cM}^{(1)}, \, \mb \Delta_C^{(1)}\right] \,,
\end{align}
up to terms resulting from the one-loop CFV effect. Here $\mb I_{\overline \cM}^{(1)}$ is the color operator that captures IR poles at one-loop in the reduced amplitude $|\cM(\tilde P, p_3,\ldots)\rangle$.
The two-loop pieces common to $\cN = 4$ and QCD read~\cite{Henn:2024qjq,Buccioni:2026mfg}
\begin{align}\label{eq:delta2c}
&\mb \Delta^{(2)}_C = 2\pi \im \, \bar\mu^{2\eps} \sum_{k>3}^n \left[\mb T_2 \cdot\mb T_{\rm in} , \mb T_2 \cdot \mb T_k\right] \bigg\{\brac{\frac{1}{\eps^2} - \frac{\pi^2}{6}}\times
\\
& \left[ (y_2 - y_k)+ \frac{\im \pi}{2} \right] +\frac{1}{6}\brac{\ln^3 e^{2\im\phi_{k2}} +4\pi^2 \ln e^{2\im\phi_{k2}} } +2\zeta_3\bigg \} \nn\, .
\end{align}
In QCD there are also finite remainder CFV pieces~\cite{Buccioni:2026mfg}.
Here we have chosen a frame where $p_1$ and $p_3$ collide head on: $p_1^\mu = -\omega n^\mu/2$, $p_3^\mu = -\omega_{\rm in} n_3^\mu/2$, with $n^2 = n_3^2 = 0$, $n\cdot n_3 = 2$, $\omega,\omega_{\rm in} > 0$, $\bar \mu = \mu^2/\mb p_{2\perp}^2$, and $\phi_{k2} \equiv \phi_k- \phi_2$ is the difference of azimuthal angles in the transverse plane.
The rapidities are defined as
\begin{align}\label{eq:y2yI}
&y_i \equiv\frac{1}{2} \ln \frac{p_i \cdot n}{p_i\cdot n_3} \, ,&
&y_2 - y_k = \frac{1}{2}\ln\brac{\frac{s_{12}s_{3k}}{s_{23}s_{1k}}} \, .&
\end{align}
In the limit $p_1|| p_2$, $y_2 \ra -\infty$. We will see below that this is a large rapidity logarithm also from the perspective of soft and collinear EFT modes having the same virtuality.

\sectionPaper{The Glauber operators}
A crucial step towards connecting the above findings with an EFT perspective was taken in \Refcite{Schwartz:2017nmr} where they exploited soft collinear effective theory (SCET)~\cite{Bauer:2000ew,Bauer:2000yr,Bauer:2001ct,Bauer:2001yt,Bauer:2002nz} with Glauber operators~\cite{Rothstein:2016bsq},
and reproduced the one-loop Glauber phase in \eq{sp1cfv} and the imaginary part of the logarithm (the $\sim \pi^2/\eps^2$) in \eq{delta2c}.
We briefly review this setup now. The operators mediating the hard scattering can be written as
\begin{align}\label{eq:hard_ops}
\mb O^{(m)}_H (x)= \! \sum_{\{\tilde n_i\}} \prod_{i = 1}^m \Phi_{\tilde n_i}^{b_i} \mb S_{\tilde n_i} (x) \: \Big |C_m\brac{\{p_i\cdot p_j\}}\Big] \,,
\end{align}
such that the amplitude at leading power is given by
\begin{align}\label{eq:cMOH}
\big | \cM \big ] =\!\! \sum_{m\leq N} \left \langle p_2p_4\ldots p_N \left | {\rm T}\left \{ \mb O^{(m)}_{H}(0) \, e^{\im S_G}\right \} \right | p_1p_3 \right \rangle \, .
\end{align}
The $\Phi_{n_i}^{b_i}$ are gauge invariant collinear building blocks with $b_i = \{ q,\bar q,g\}$ for radiation collinear
to the light-like vectors $n_i$ along $p_i$.
The Wilson lines $\mb S_{n_i}$ source soft gluons and act as matrices in the color space, modifying the contraction of external collinear legs with the Wilson coefficients $|C_m]$ (see \Refcite{Pathak:2021wdr}).
The physics associated with factorization violation is described by the Glauber action $S_G = \sum_{n_i} S_G^{(n_is)} +\sum_{(n_i,n_j) }S_{G}^{(n_isn_j)}$ \cite{Rothstein:2016bsq}, with
\begin{align}\label{eq:SG}
S_G^{(n_isn_j)} &= 8\pi \as \sum_{ab} \int \df^d x \df^d y \df^d z \int \frac{\df^d q_i \df^dq_j}{(2\pi)^{2d}} \\
&\hspace{-30pt}\times \frac{e^{\im q_i\cdot(x-z ) + \im q_j \cdot (y-z)}}{q_{i\perp}^2q_{j\perp}^2}
\mb \cO_{n_i,x}^a \cdot \hat{\mb \cO}_{s,z}^{(n_in_j)} (q_{i\perp},q_{j\perp}) \cdot \mb \cO_{n_j,y}^b \,, \nn\\
S_G^{(n_is)} &= 8\pi \as \sum_{ab}\! \int\! \frac{\df^d x \df^d y\df^d q_i }{(2\pi)^d}\frac{e^{\im q_i\cdot (x-y)}}{q_{i\perp}^2} \mb \cO_{n_i,x}^a \cdot \mb \cO_{s,y}^{b_{n_i}} \nn \, .
\end{align}
The operators $\mb{\cO}_{n_i,x}^{a}$ etc. are gauge invariant bilinears of collinear and soft fields at location $x$, and $a,b = q,g$.
The collinear momenta scale as $p_i = (n_i\cdot p_i, \bn_i \cdot p_i, p_{i\perp}) \sim Q(\lambda^2,1,\lambda)$ for a power counting parameter $\lambda \ll 1$ and $\bn_i$ is an auxiliary vector used to define the light cone decomposition relative to $n_i$. On the other hand, the soft momenta scale as $k_s \sim Q(\lambda,\lambda,\lambda)$. The scaling of the offshell Glauber momenta $q_i$ is discussed below.

The Glauber-SCET formalism has been used to derive factorization in the forward scattering limit such as the small-$x$ limit of DIS~\cite{Neill:2023jcd}, diffraction~\cite{Lee:2025fml,Aretz:2026sjb}, saturation~\cite{Stewart:2023lwz} and the Regge factorization of amplitudes~\cite{Moult:2022lfy,Rothstein:2023dgb,Rothstein:2024fpx,Gao:2024qsg,Gao:2024fyz}.
Here the large logarithms $\sim \ln \abs{s/t}$, where $t\ll s$ is the momentum transfer, translate into \textit{rapidity divergences} in the collinear and soft matrix elements of Glauber operators, and the exponentiation in these logarithms is achieved through a ``rapidity renormalization''.
This is organized in the number of insertions of the Glauber actions, $S_G$, where each insertion adds another order in the logarithmic resummation counting.
Note that the kinematics in the forward scattering directly constrains the momentum transfer to follow the Glauber scaling. Instead, in this \letter we derive for the first time a factorization involving Glauber gluons exchanged \textit{alongside a hard scattering} where the momentum transfer is constrained to be Glauber-like not by external kinematics, but \textit{via pinching} in the loop momentum space.

\sectionPaper{Factorization of subgraphs}
In \eq{SG}, it is assumed that the Glauber momentum $q_{i}$ \textit{injected} into the $n_i$ collinear sector satisfies the power counting $q_i^+ = n_i\cdot q_i \sim Q\lambda^2$, $q_{i\perp}\sim Q\lambda$, with $q_i^-=\bn_i \cdot q_i \lesssim Q\lambda$. However, it is often possible to deform the integration over $q_i^+$ away from the Glauber region to $q_i^+ \gtrsim Q\lambda$ in the complex plane. In the original formulation in \Refcite{Rothstein:2016bsq} this scaling is enforced regardless and subtractions are required everywhere.
While this true to the spirit of an EFT description, it creates a major hurdle in factorizing soft and collinear subgraphs. This is because enforcing Glauber scaling when the momentum is \textit{not} pinched requires a non-analytic regulator of the form $\abs{\brac{q_i^+-q_i^-}/\nu}^{-\eta} $ to create an \textit{artificial pinch}. However, such a regulator entangles the two longitudinal components and makes it impossible to perform their multipole expansion, the backbone of deriving factorization theorems in QCD. Instead, we would like $q_i^+\sim Q\lambda^2$ to be dropped in the soft matrix elements with momenta $ \ell_s^+ \sim Q\lambda$, and the same for $q_i^-$ in the collinear matrix elements characterized by the large $p_i^- \sim Q$ momentum carried by the external legs.

\begin{figure}[t]
\centering
\includegraphics[width=0.45\textwidth]{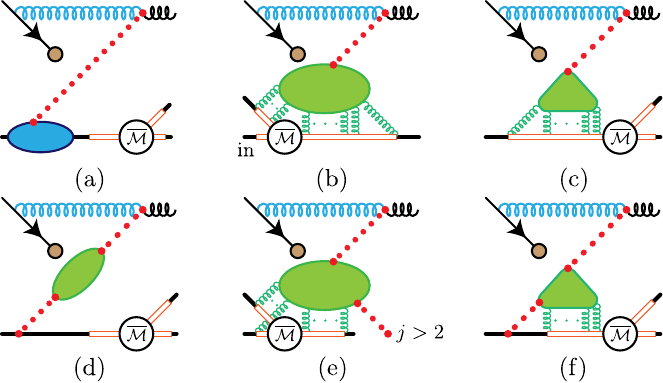}
\caption{Topologies vanish either upon contour integration of longitudinal momenta (a, b, c, d), cancel against the subtraction terms (e), or are scaleless in rapidity (f).}
\label{fig:blobs}
\end{figure}

We overcome this problem by 1) adopting the strategy of deforming the contour wherever possible
and employing the Glauber operators in the CFV pieces only when $q_i^+$ \textit{remain pinched} and 2) employing a new regularization scheme for the \textit{true rapidity divergences} that does not spoil analyticity in the soft, collinear and Glauber longitudinal momenta.
As discussed below, this approach drastically minimizes the clutter and enables us to successfully factorize the CFV pieces. For a single gluon pinched in the Glauber region, we have derived
\begin{align}\label{eq:fact}
&\overline \cM^{[0]}(\tilde P)\mb \Delta_{C}^{\text{cfv}} \spl^{(0)} (p_2)= \frac{2 \pi\im}{4\pi} \int \dfbar^{d-2} q_\perp \int\dfbar^{d-2} q_{\perp}^\prime \\
& \quad \times \frac{1}{\mb q_\perp^2}\mb C \brac{p_2, q_\perp, \frac{\nu^-}{\omega}; p_1} \cdot \widehat{\mb S} \brac{q_\perp, q_{\perp}',\nu^+ ; n_k } \cdot \frac{\mb T_{\rm in}}{\mb q_{\perp}^{\prime2}} \, ,\nn
\end{align}
where $\dfbar\! \equiv\! \df/(2\pi)$. The reduced amplitude $\overline \cM= \langle \tilde P| \Phi_n^b(0) | 0\rangle$, with $b \in \{q,\bar q, g\}$, needs to be evaluated only at the tree-level, as the factorization captures solely the new CFV pieces.
We see that \eq{fact} has a remarkably constrained form: a) it excludes any possibility of a non-trivial $n_j$-collinear subgraph for $j\neq 1,2$, b) while the
soft subgraph in \eq{fact} induces correlations with other legs $n_k$, it necessarily connects to $p_3$ via a Glauber propagator $1/\mb q_\perp^{\prime 2}$, and c) there are no Glauber propagators associated with final state legs.
This requires that graphs with topologies shown in \fig{blobs} that do not confirm to the picture in \fig{fact} vanish to all loops.
We have sketched the derivation of \eq{fact} in the \textit{End Matter}.

\sectionPaper{The collinear subgraph}
The crux of the factorization is in the definition of the collinear subgraph:
\begin{align}\label{eq:coll_def}
&\mb C\brac{ p_2, q_\perp, \frac{\nu^-}{\omega}; p_1} \equiv \sum_a \int \dfbar q^+ \int \df^d x \: e^{\im \frac{q^+ x^-}{2} +\im q_\perp\cdot x_\perp} \nn \\
&\qquad \qquad\quad \times \br{p_2} {\rm T} \left\{\mb \cO_n^a (x) \Phi^b_n (0)
\right\}\kt{p_1}_{\text{deform $q^+$}}\, ,
\end{align}
where $q^+ = n\cdot q$. Here, we require that all the graphs that feature \textit{single $q^+$ propagators} of the form $[- q^+ +\Delta+ \im 0]^{-1}$, where $\Delta\sim Q\lambda^2$, are deformed into the soft region $q^+\ra \lambda Q$. These correspond to the $\mb \cO_{n}^b$ operator in \eq{SG} Wick-contracted with the building block $\Phi_n^b$ in \eq{hard_ops} moving the Glauber loop entirely into the reduced amplitude. See the \textit{End Matter} for the details. However, for space-like kinematics there are cases where $q^+$ is pinched to $\sim Q\lambda^2$, and only these make a nontrivial contribution to $\mb C$.
This can never happen for the subgraphs in \fig{blobs}(a) associated with other legs ($j > 2$).

At the tree-level, the pinching in $q^+$ results from the Glauber insertion on the outgoing parton, $p_2$. For $b = q$:
\begin{align}\label{eq:C0}
\mb C^{[0]} = \includegraphics[height=1.3cm,valign=c]{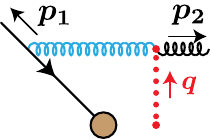}
= \mb T_2
\spl^{(0)} \!\brac{p_2-q}
\frac{\nslash\bnslash}{4} u (p_1)
\, .
\end{align}
Combining \eq{C0} with the tree-level soft function,
\begin{align}\label{eq:S0}
\left[A\left|\widehat{\mb S}\right|B\right]^{[0]} \!= 2g_s^2 \delta^{AB}\mb q_\perp^2 (2\pi)^{d-2} \delta^{(d-2)} (q_\perp\! +\! q_{\perp}') \, , \!
\end{align}
reproduces the one-loop result in \eq{sp1cfv}~\cite{Schwartz:2017nmr}.
The soft subgraph construction will be explained below.

We now compute the collinear subgraph to one-loop for the two-loop CFV.
It should be noted that Glauber-SCET matrix elements with rapidity divergences cannot be extracted from QCD amplitudes, which invalidates the use of standard tools. We carried out computations in the \texttt{SCETCalc} framework~\cite{scetcalc} developed by some of us, interfacing \texttt{QGRAF}~\cite{Nogueira:1991ex}, \texttt{FeynCalc}~\cite{Mertig:1990an,Shtabovenko:2016sxi,Shtabovenko:2020gxv}, \texttt{FORM}~\cite{Ruijl:2017dtg}, and Pak's cut-filter algorithm~\cite{Pak:2011xt} to automate the fixed order calculations directly using SCET Feynman rules.
We employed a $\abs{\cP^-/\nu^-}^{-\eta}$ regulator~\cite{Chiu:2012ir,Neill:2023jcd} that allows the $\ell^+$ components of the collinear loop momenta to be integrated via contours.
At one-loop for different channels we find $\sim 130$ graphs. However, after deforming the single $q^+$ propagator ones into the soft region, only $\sim 15$ survive upon performing contour integration. In other words, only in these graphs is $q^+$ pinched to $\sim Q\lambda^2$.

Remarkably, of the remaining pinched graphs, we find that only 5 exhibit rapidity divergences.
The only sources of the rapidity poles are $n$-collinear Wilson lines inside the Glauber operators and the building block $\Phi_n^b$.
However, rapidity poles in the following two graphs cancel:
\begin{align}
\includegraphics[height=1.3cm,valign=c]{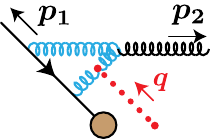}
+
\includegraphics[height=1.3cm,valign=c]{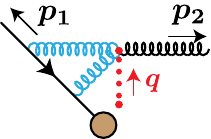} = \text{rapidity finite} \,.
\end{align}
The rapidity divergence instead arises from two graphs involving emission from the Wilson line inside $\Phi_n^q$:
\begin{align}\label{eq:coll_real}
& \includegraphics[height=1.3cm,valign=c]{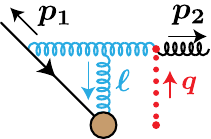}
+ \includegraphics[height=1.3cm,valign=c]{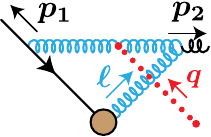} = -\as \brac{\frac{1}{\eta} + \ln \frac{\nu^-}{\omega}} \\
&\qquad \qquad \qquad \times \mb C^{[0]} \: \mb T_2 \mb T_1 \omega_G(p_{2\perp} -q_\perp) \brac{1 + \cO(\eta)} \, ,
\nn
\end{align}
and \textit{the Regge pole}~\cite{Rothstein:2016bsq,Moult:2022lfy,Gao:2024qsg} (up to $\cO(\eta^0)$ terms):
\begin{align}\label{eq:coll_regge}
\includegraphics[height=1.3cm,valign=c]{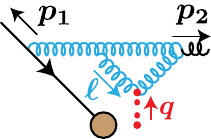} = \as N_c \omega_G(q_\perp) \brac{\frac{1}{\eta} + \ln \frac{\nu^-}{\omega}} \mb C^{[0]} \, ,
\end{align}
Here, $\omega_G$ is the leading order Regge trajectory:
\begin{align}
\omega_G(q_\perp) &\equiv \brac{\frac{\mu^2 e^{\gamma_E}}{4\pi}}^{\eps}\int \dfbar^{d-2} \ell_\perp \: \frac{\mb q_\perp^2}{\mb \ell_\perp^2 \brac{\mb \ell_\perp -\mb q_\perp}^2} \\
& = \frac{e^{\gamma \epsilon } \Gamma (-\epsilon )^2 \Gamma (\epsilon +1) }{4 \pi \Gamma (-2 \epsilon )}\left(\frac{\mb q_{\perp}^2}{\mu ^2}\right)^{-\epsilon } \nn \, .
\end{align}
Thus, we have arrived at a simple result:
\begin{align}\label{eq:C1bare}
\mb C^{[1] \: \text{bare}} &= - \as\brac{\frac{1}{\eta} + \ln\frac{\nu^-}{\omega} + \cO(\eta^0)} \\
&\quad \times \mb C^{[0]}\brac{\mb T_2 \mb T_1 \omega_G (p_{2\perp} -q_\perp) - N_c \omega_G(q_\perp)} \,. \nn
\end{align}
Notice that the result involves contributions not only from the hard Wilson lines in the first term but also the Regge pole, the hallmark of forward scattering.

Combining \eq{C1bare} with \eqs{fact}{S0}, we find
\begin{align}\label{eq:C1S0}
&\mb C^{[1]\: \text{bare}} \otimes \widehat{\mb S}^{[0]} :\brac{\frac{\as}{2\pi}}^2 2 \pi\im \bar \mu^{2\eps} \brac{\frac{1}{\eps^2} - \frac{\pi^2}{6} + \cO(\eps)}\\
&\quad\!\times\brac{\frac{1}{\eta} + \ln\frac{\nu^-}{\omega}}\!\left[- \mb T_2 \spl^{(0)}(p_2) \brac{\mb T_2 \mb T_1 - \frac{N_c}{2}}\right]\cdot \mb T_{\rm in}\nn \,.
\end{align}
As expected, this is independent of the collinear parton flavors and we confirmed this by computing all the cases.

\sectionPaper{The soft subgraph}
Now, consistently canceling the $1/\eta$ above against the soft subgraph
requires a BFKL-ladder like color structure.
To see this, let us relate the anticipated color structure in \eq{delta2c} with \eq{C1S0}:
\begin{align}\label{eq:cfv2col}
&\sum_{k>3} \left[A_2 \left|\left[\mb T_2 \cdot \mb T_{\rm in} , \mb T_2 \cdot \mb T_k \right]\right|B_2\right] \mb T_1^{B_2} y_2 \\
&= \! \sum_{k>3} y_2\left[A_2 \left|\mb T_2^{A_{\rm in}}\right|B_2\right] \mb T_1^{B_2}\: \nn
\left[A_{\rm in} \left|\mb T_2^{B_{\rm in}}\right| C_k\right]\mb T_k^{C_k}\mb T_{\rm in}^{B_{\rm in}} \\
&= -y_2 \left[A_2 \left|\brac{\mb T_2 \mb T_1} \brac{\mb T_2 \mb T_1 - \frac{N_c}{2}}\right|B_{\rm in}\right] \mb T_{\rm in}^{B_{\rm in}}
\, ,
\nn
\end{align}
where we isolated the $y_2$-term, and $[A | \mb T_2^C | B] = \im f^{ACB}$.
The second line can be schematically represented as
\begin{align}\label{eq:cfv2col_schematic}
\includegraphics[height=2cm,valign=c]{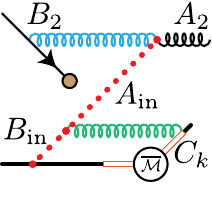} \, .
\end{align}
With $\spl^{(0)} \propto \mb T_1$, we recognize the upper part of \eq{cfv2col_schematic}
as the tree-level collinear subgraph $\mb C^{[0]} \propto \mb T_2 \mb T_1$. In the third line, we find a result consistent with \eq{C1S0} after applying color conservation for the $y_2$-term.
We will now show that the one-loop soft subgraph indeed reproduces the required color structure in \eq{cfv2col}.

Interestingly, given the complete absence of collinear dynamics along the directions $n_j$, $j\neq 1,2$, we can treat the rapidity divergences in the soft sector analogous to the collinear case via a new $\left| \cP^+/\nu^+\right|^{+\eta/2}$ regulator where the divergence for $\ell^+\ra0$ is regulated for $\eta > 0$. Note that, in contrast with the $\nu^-$ employed in the $n$-collinear sector, the $\nu^+$ tracks the $+$-momentum, with the combination $\nu^+\nu^-\sim \mb q_\perp^2 \sim \mb p_{2\perp}^2$ being rapidity invariant.
Unlike the prescription of $|2\cP^z/\nu|^{-\eta/2}$-type regulators in \Refcite{Chiu:2012ir,Rothstein:2016bsq,Moult:2022lfy,Gao:2024qsg} forcing upon us significantly complicated integrals of soft loop $\ell_s^0$ with quadratic propagators, this regulator offers the unique advantage of maintaining analyticity in the $\ell_s^-$ and $q^-$ momenta, enabling their integration via contours and preserving their multipole expansion in the collinear subgraph, and that of $q^+\sim \lambda^2 Q$ in the soft subgraphs.
Consequently, we find similar dramatic simplifications also in the soft sector: the cases (a), (b), (c), (d) in \fig{blobs} yield vanishing contour integrals at all-loops~\cite{cfv_long}!
As explained in the \textit{End Matter}, we expect the graphs of type (e) to cancel against the zero-bins of graphs without the bottom Glauber rung to all-loops, whereas the graphs of type (f) with our regulator becomes scaleless in rapidity.
Hence, similar to the collinear sector, only those graphs contribute where all the soft $\ell_s^-$ and $q^-$ momenta get pinched, leading to the following operator definition of the soft subgraph:
\begin{align}\label{eq:soft_def}
&\int \dfbar^{d-2}q_\perp' \: \widehat {\mb S} \brac{q_\perp, q_\perp' } \cdot \frac{\mb T_{\rm in}}{\mb q_\perp^{\prime 2}} = \sum_b \int \frac{\dfbar q^-}{2} \int \df^d y \\
\nn &\quad \times
e^{-\im \frac{q^- y^+}{2}-\im q_\perp \cdot y_{\perp}}
\: \br{0}{\rm T} \left\{\mb \cO_{s}^{b_{n}}\brac{y} \prod_{j=3}^{n} \mb S_{n_j} (0) \right\}\kt{0}_{\nu^+}
\end{align}

At one-loop, only correlations between the incoming and an outgoing Wilson lines are pinched, such that
\begin{align}\label{eq:S1}
&\widehat{ \mb S}^{[1]\text{bare}} =\sum_{k>3} \includegraphics[height=1.6cm,valign=c]{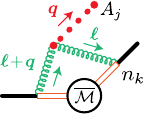}= -16\pi\as^2 \nu^\eta \int_0^\infty \! \frac{\df \ell^+}{\brac{\ell^{+}}^{1-\eta}}
\nn \\
&\times\brac{\frac{\mu^2e^{\gamma_E}}{4\pi}}^{\eps}\sum_{k>3} \mb T_2 \mb T_k \frac{\brac{\mb q_\perp + \mb n_{k\perp}\ell^+/n_k^+}^2 }{\brac{\mb q_\perp+ \mb q_\perp'+ \mb n_{k\perp}\ell^+/n_k^+}^2} \, ,
\end{align}
This graph at first sight does not seem to fit the generic scenario in \fig{fact} but upon contour integration we find $\ell^- + q^-\ra 0$, turning $[\brac{\ell+q}^2+\im0]^{-1}$ effectively into a Glauber propagator with $q_\perp' = -\ell_\perp -q_\perp$.
Note that, it is in this sense we interpret the bottom red-dotted propagator in \fig{fact} to be Glauber-like, and not necessarily as an explicit insertion of Glauber operator on the $j = 3$ line. The latter is contained in \fig{blobs} (e) and (f), and discussed further in the \textit{End Matter}.
Next, we have
\begin{align}\label{eq:soft_rescatter_jk}
&\int \frac{\dfbar^{d-2}q_{\perp}'}{\mb q_\perp^{\prime 2}}\widehat{\mb S}^{[1]\text{bare}}\brac{q_\perp, q_\perp'} \cdot \mb T_{\rm in}= \! -\frac{16\pi \as^2}{\eta}\omega_G(q_\perp)\!\\
\nn
&\quad \times \sum_{k>3}\brac{\mb T_2 \cdot \mb T_{\rm in}} \mb T_k\nn
\brac{\frac{\nu^{+2} \mb n_{k\perp}^2}{n_k^{+2}\mb q_\perp^2}}^{-\eta/2} \brac{1 + \cO\brac{\eta^0}} \, ,
\end{align}
reproducing the desired color structure in \eq{cfv2col}.
Combining with $\mb C^{[0]}$ in \eq{C0} following \eq{fact}, and setting $\nu^+\nu^- \ra \mb p_{2\perp}^2$, the rapidity pole in \eq{C1S0} is canceled, reproducing the term $\propto (y_2-y_k)$ in \eq{delta2c}. The remaining terms in \eq{delta2c} can be computed from the rapidity finite pieces in \eq{soft_rescatter_jk}.
Finally, the finite remainder CFV pieces computed in \Refcite{Buccioni:2026mfg} that do not appear in the $\cN=4$ result in \eq{delta2c} will arise from the rapidity finite terms in the collinear subgraphs. The computation of these terms will be presented in \Refcite{cfv_long}.

The soft subgraph in \eq{soft_rescatter_jk} in combination with the collinear subgraph constitutes the simplest example of hidden (Glauber) regions in amplitudes.
Interestingly, as discussed in the \textit{End Matter}, the rapidity divergences in the soft subgraphs cancel against those in \eq{C1S0}, which, on the contrary, are leftover pure Glauber phases resulting from the differing $\im0$ prescription in the incoming and outgoing Wilson lines. Hence, we see that the CFV pieces arise from \textit{both} Glauber phases and pinched Glauber regions, which is why we have taken special care in defining the tree-level objects in \eqs{C0}{S0}. This is expected since the large rapidity logarithm $y_2$ cannot arise from the soft sector alone.

We expect that similar to the collinear subgraph story above, only a small fraction of higher loop soft subgraphs (apart from the vanishing cases in \fig{blobs}) will survive after $\ell_s^-$ and $q^-$ contour integrations.
Thus,
the procedure of contour integrals of the appropriate longitudinal momenta in the EFT subgraphs
can be turned into a new algorithm for isolating hidden regions at any loop, and it would be interesting to compare our approach with \Refcite{Gardi:2024axt}. Finally, we do not see any conceptual difficulty in generalizing this approach to the case of multiple Glauber rungs in \fig{fact}.

\sectionPaper{Reggeization in factorization violation}
A key feature of Glauber SCET is clean separation of soft and collinear matrix elements in the rapidity, enabling their all order resummation, such as in \Refscite{Neill:2023jcd,Lee:2025fml,Moult:2022lfy,Gao:2024qsg,Gao:2024fyz}.
However,
here $q_\perp,q_\perp'$ are not constrained by external kinematics and there is some freedom in reshuffling them under the integral. Based on our one-loop data, it is unclear if the rapidity evolution equation is a BFKL-like convolution or a multiplicative one like the Regge trajectory.
To this end, we follow the approach of \Refcite{Catani:2011st} where the IR factorization formula was used to fix the IR poles in $\spl$:
\begin{align}\label{eq:cdfr}
\mb \spl = \left[1- \overline {\mb V} (\eps){\mb V}^{-1}(\eps)\right]\spl + \overline {\mb V}(\eps) \spl_{\text{fin.}}\overline {\mb V}^{-1}(\eps)\,,
\end{align}
where $\mb V^{-1} (\eps)\equiv 1 - \mb I_\cM(\eps)$, and likewise $\overline{\mb V}(\eps)$ for $\overline \cM$.

We focus on the leading pole terms with the highest power of the rapidity logarithm, $y_2$, that scale as $\as^\ell /\eps^{\ell}y_2^{\ell -1}$.
These arise from differences of the IR operators $\mb I_{\cM}$ and $\mb I_{\overline \cM}$ acting on the tree-level $\spl^{(0)}$ in \eq{cdfr}, whereas terms involving $\spl_{\text{fin}}^{(i>0)}$ do not lead to the highest pole.
As seen in the two-loop result in \eq{sp2cfv}, the appearance of $\mb I_{\overline \cM}^{(1)}$ is essential to involve the kinematic invariants associated with the non-collinear partons,
\begin{align}
\mb I_{\overline \cM}^{(1)} = \frac{-1}{2\eps}\sum_{j,k>2} \mb T_j\cdot \mb T_k \brac{L_{jk} - L_{j\tilde P} - L_{k\tilde P}} + \ldots \,,
\end{align}
where $L_{ab}\equiv \ln[ (-s_{ab} -\im 0)/\mu^2]$, and the other terms vanish in the commutator with the Glauber phase in \eq{sp1cfv}. Careful inspection of the three-loop terms $\as^3/\eps^3y_2^2$ reveals that the $\ell$-loop $\as^\ell /\eps^{\ell}y_2^{\ell -1}$ term arises from the commutator of $\mb I_{\overline \cM}^{(1)}$ with the $(\ell\!-\!1)$-loop rapidity log. This is captured by the following leading logarithmic (LL) result:
\begin{align}\label{eq:deltacLL}
\mb \Delta_{C,\text{LL}}^{\text{cfv}} = \sum_{\ell = 0}^\infty \frac{2\pi \im(2y_2)^\ell}{(\ell+1)!}\!\brac{ \frac{-\as}{2\pi\eps}}^{\!\!\ell+1} \!\! \text{ad}_{\mb T_2 \cdot \mb T_1}^{\ell}\!\! \brac{\mb T_2 \cdot \mb T_{\rm in}} ,\!\!
\end{align}
where the color operator is given by nested commutators:
\begin{align}
\text{ad}_{\mb T_2 \cdot \mb T_1}^{\ell}\!\! \brac{\mb T_2 \cdot \mb T_{\rm in}} &\equiv \underbrace{\big[ \mb T_2 \cdot \mb T_1 ,\, \big[ \ldots \big[ \mb T_2 \cdot \mb T_1}_\text{$\ell$ times} ,\mb T_2\cdot \mb T_{\rm in} \big] \big]\nn \\
&= \brac{\mb T_2 \mb T_1 - \frac{N_c}{2}}^{\ell} \mb T_2 \cdot \mb T_{\rm in}\, ,
\end{align}
corresponding precisely to the ladder structure in \eq{cfv2col_schematic}.
The result in \eq{deltacLL} matches our explicit computation of the $\as^3/\eps^3y_2^2$ term and results from
a multiplicative evolution derived from \eqs{C1bare}{soft_rescatter_jk}:
\begin{align}\label{eq:RRGE}
\nu^+\frac{\partial}{\partial \nu^+} \widehat{\mb S} \brac{q_\perp, q_\perp' , \nu^+ } &= \mb \gamma_\nu (q_\perp)\, \widehat{\mb S} (q_\perp, q_\perp',\nu^+) \,,\\
\nu^-\frac{\partial}{\partial \nu^-} \widehat{\mb C} \brac{p_2, q_\perp, \nu^-}&= \mb C\brac{p_2, q_\perp, \nu^-} \mb \gamma_\nu (q_\perp) \,,\nn
\end{align}
with the one-loop rapidity anomalous dimension given by
\begin{align}\label{eq:gammanu}
\frac{\as}{2\pi} \mb \gamma^{[1]}_\nu (q_\perp)= -\as \brac{2\mb T_2 \cdot \mb T_1 - N_c} \omega_G(q_\perp) \, .
\end{align}
Note that for the $\mb T_2 \mb T_1$ term in the collinear function, we had to shift the variable $q_\perp \ra p_{2\perp} - q_{\perp}$ in \eq{coll_real} under the $q_\perp$ integration. It would be interesting to compute $\mb C$ to one higher order to better understand this point. Combining the LL solution of \eq{RRGE} with \eq{fact} yields
\begin{align}\label{eq:factLL}
\mb \Delta_{C,\text{LL}}^{\text{cfv}} =\frac{4\im \pi\as }{ \overline \cM^{[0]} \spl^{(0)}} \int \frac{\dfbar^{d-2}q_\perp}{\mb q_\perp^2} \mb C^{[0]} e^{\mb \gamma_\nu (q_\perp) y_2 } \cdot \mb T_{\rm in} \, ,
\end{align}
where every term in the series is a simple bubble integral with the leading pole given by \eq{deltacLL}.

Thus, we are now able to endow a definite meaning to the results in \eqs{delta2c}{deltacLL} beyond a consequence of laborious manipulations of terms in the IR factorization formula. It is apparent from the graphs in \eq{coll_real} contributing to the $\mb T_2\mb T_1$ term, and likewise from \eq{S1}, that the usual soft and collinear Wilson lines in the hard operator are at play in generating these rapidity poles. Interestingly, this is not the complete story!
The Regge trajectory, the $N_c$ part of $\mb \gamma_\nu$ in \eq{gammanu}, is hiding inside the results in \eqs{delta2c}{deltacLL}, and the $e^{\mb \gamma_\nu (q_\perp) y_2 }$ in \eq{factLL} ``reggeizes'' the exchanged gluon.
It is interesting to note that
the rapidity divergences at each order result from a mechanism similar to the ``breaking of the plus prescription'' identified in \Refcite{Forshaw:2006fk} in the context of superleading logarithms (SLLs) (see also \Refcite{Becher:2024kmk}).
Finally, hints of a BFKL ladder-like color structure in the CFV pieces associated with SLLs were observed in \Refcite{Forshaw:2008cq}.
In this work, we have made this connection precise.

\sectionPaper{Conclusions}
In this \letter we exploited the Glauber operators of \Refcite{Rothstein:2016bsq} to shed light on the all-order IR structure of collinear factorization violating pieces. We overcame a number of problems associated with the non-analytic regulator in the Glauber SCET Lagrangian for the collinear factorization violation problem by combining ideas of contour deformation and designing a new rapidity regulator that maintains analyticity in the longitudinal momenta and unlocks their multipole expansion in EFT subgraphs.
Aided by the development of a new software tool \texttt{SCETCalc}~\cite{scetcalc}, we demonstrated a dramatic simplification arising from the pole structure in the loop momentum space in the EFT subgraphs. As a consequence, we found entire topologies of graphs vanishing to all loops, leading to the all-order picture in \fig{fact}.
We also identified the origin of the various finite CFV pieces in terms of the collinear and soft subgraphs which yields a significant computational advantage compared to the full multi-point amplitude.

With the clean separation of the IR modes in the rapidity, we explained the physical origin and the exponentiation of the large rapidity logarithm observed by \Refscite{Catani:2011st,Forshaw:2012bi}.
The appearance of Regge trajectory reveals an interesting secret interplay of forward scattering dynamics alongside a hard scattering.
The advancements achieved here lay the foundation for tackling such phenomena at the cross section level in non-global observables at hadron colliders and systematically investigating the factorization of the IR effects into PDFs.

\sectionPaper{Note Added}
Near the completion of this paper we have been made aware of a related work by Chen, Gardi, Ma, Ma, Zhang, and Zhu~\cite{cgmmzz}. Both have been presented at the Parton Shower and Resummation conference at the University of Manchester, 8-10 July 2026.

\begin{acknowledgments}
\headingAcknowledgments
We thank Yao Ma, Iain Stewart, Einan Gardi, Jeffery Forshaw, Christoph Dlapa, Aniruddha Venkata, and Federico Buccioni for helpful discussions.
DB and AP are funded by the European Union (ERC, TOPMASS, 101165601). Views and opinions expressed are however those of the author(s) only and do not necessarily reflect those of the European Union or the European Research Council. Neither the European Union nor the granting authority can be held responsible for them.
AG is supported by the United States Department of Energy under contract DE-AC02-76SF00515.
Part of the work of JRG has been supported by the Royal Society through Grant URF\textbackslash{}R1\textbackslash{}201500.
\end{acknowledgments}

\bibliography{qcd}

\clearpage

\section{End Matter}

\begin{figure}[b]
\centering
\includegraphics[width=0.42\textwidth]{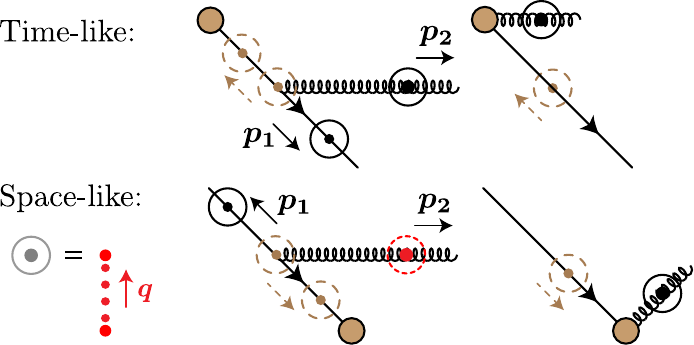}
\caption{Glauber operators inserted inside solid black circles vanish, and those in dashed brown can be deformed into the reduced amplitude.
Only in the space-like splitting is $n\cdot q$ pinched to $\sim Q\lambda^2$ for the insertion in the dotted red circle.}
\label{fig:tree_insertions}
\end{figure}

\sectionPaper{Derivation of the factorization}
In this appendix, we sketch the main elements of derivation of the factorization presented in \eq{fact}.
Let us begin with considerations of collinear loops and throw in a single insertion of the Glauber operator.
In the collinear limit, we consider $m = N-1$ in \eq{cMOH}, such that $p_{1}$ and $p_2$ are obtained from a single collinear matrix element $\br{p_2} \Phi_{n_1}^b \kt{p_1}$. Note that the $\tilde n_i$ in \eq{hard_ops} is integrated over, and is set to an appropriate direction depending on the state. We will write $n_1 = n$.
We then consider time-ordered products
$\br{p_2} {\rm T} \, \Phi_{n}^b (0)\mb \cO_{n}^a(x)\kt{p_1}$
that begin to tie the $n$-collinear matrix elements to other sectors. However, there are a number of simplifications resulting from being able to deform the ``un-pinched'' Glauber momenta.
To illustrate the idea, let us consider all possible Glauber operator insertions in the tree-level amplitude for time-like and space-like splittings for $b = q$ shown in \fig{tree_insertions} with the operator $\mb \cO_{n}^a$ injecting momentum $q$.

Firstly, an insertion contracted with longitudinally polarized $\bn \cdot A_n$ Wilson line emissions vanishes as the $\mb \cO_n^g$ involves $\cB_{n\perp}^\mu$ building blocks with $\bn \cdot \cB_{n\perp} = 0$.
Next, we have the other insertions shown in solid black circles.
Note the $q^- \equiv \bn \cdot q \lesssim Q\lambda$ component can be multipole expanded away relative to $p_{1,2}^-\sim Q$. As a result, it is straightforward to check that the graphs corresponding to black-circle insertions involve two propagators with poles in $q^+ \equiv n\cdot q$ momentum on the same side of the real axis.
We can then pick the pole in $q^0 \equiv (q^++q^-)/2$ corresponding to the lower subgraph in \fig{fact}, and find all the poles in $q^z \equiv(q^- - q^+)/2$ on the same side of the real axis, leading to a vanishing result. Note that the lower part may feature a soft subgraph with a convergent $q^-$ integration (see \eq{S1}), in which case we can simply integrate over $q^+$ independently.
More generally, for such insertions on the \textit{active partons} entering the hard scattering, this is a consequence of the Coleman-Norton theorem~\cite{Coleman:1965xm} that requires the large momentum flow on collinear lines to be in the same direction.
It is for this reason, graphs of the topology (a) in \fig{blobs} must vanish.

\begin{figure}[t]
\centering
\includegraphics[width=0.3\textwidth]{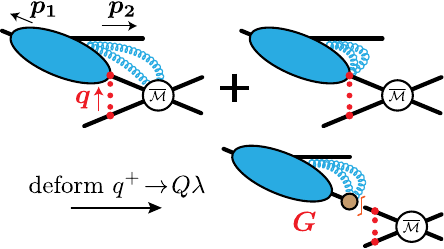}
\caption{In the graphs featuring single $q^+$ propagators, the Glauber loop can be moved into the reduced amplitude by deforming $q^+\sim Q\lambda^2 \ra Q\lambda$.}
\label{fig:all_loop_Wl}
\end{figure}

Then, there are cases shown in brown dashed circles which consist of a single propagator carrying $q^+$. Here, we note that the $q^+$ is not pinched and can be deformed away from the Glauber region, $Q\lambda^2 \ra Q\lambda$. As a result, these propagators become \textit{hard-collinear} and get contracted with the hard vertex (the brown blob).
The three graphs associated to these insertions combine for the time-like case to give
\begin{align}
\int\df^d x \: \langle qg | {\rm T} \: \mb \cO_{n_i}^{qA} (x) \: \overline \chi^\beta_{n_i} (0) | 0\rangle^{[0]}\! \ra \!\frac{- \mb T_i^A \overline u_{n_i}^\gamma \spl^{(0)}_{\gamma\beta} }{-n_i\cdot q +\im0 } ,\!
\end{align}
and analogously for the space-like case. Thus, the Glauber insertions for these cases (brown-dashed in \fig{fact}) are now moved into the reduced amplitude $\overline \cM$ and do not violate factorization.
In fact, as depicted in \fig{all_loop_Wl}, this observation concerning Glauber insertions on the parent parton $\tilde P$ generalizes to arbitrary subgraphs including virtual loops.
This relies crucially on cancellations between the Wilson line emissions of Glauber operator after $q^+$ deformation with those from the Wilson line in the collinear building block $\Phi_{n}^b$ in the hard scattering operator. Note that when we speak of ``deforming the Glauber momentum'', we do not attempt to further absorb it into the soft or collinear contributions as is done in the CSS formalism~\cite{Collins:1985ue,Collins:1988ig}. As shown in \fig{all_loop_Wl}, the final outcome is equivalent to adding a Glauber loop correction to the reduced amplitude.
Consequently, argument we have presented here does not rely on the validity of the ``Soft-Glauber correspondence'' conjectured in \Refcite{Rothstein:2016bsq}.

Finally, we see that the space-like case features one particular insertion on the outgoing parton (red-dotted circle) where the $q^+$ is indeed pinched to $\sim Q\lambda^2$ and it now gives a new contribution.
Remarkably, as noted above, once the aforementioned single $q^+$-propagators are deformed away, only a small fraction of numerous possible insertions at higher orders lead to $q^+$ being actually pinched.
The key point here is our ability to perform the $q^+$ integral independently now.
Let us consider now a generic collinear subgraph where there are at least two propagators with poles at $q^+ = \pm \delta^{(\pm)} (p_1, p_2, \{\ell_i\}) \pm \im 0$, where $\delta^{(\pm)} \sim Q\lambda^2$ can depend on arbitrary collinear loop momenta $\ell_i$. Then, for Glauber exchange with $j >2$ leg,
\begin{align}\label{eq:coll_generic}
\includegraphics[height=2.3cm,valign=c]{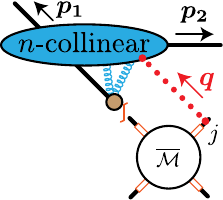} &= \im 2 g_s^2 \int \frac{\dfbar^{d-2}q_\perp}{\mb q_\perp^2} \mb\cI_n^{A_j} \brac{q_\perp,p_1, p_2}\\[-15pt]
&\quad \times \int_\text{reg} \dfbar q^z \: \frac{-\im \mb T_j^{A_j}}{2q^z \pm \delta^{(\pm)} \pm \Delta_j \pm \im 0}\nn \\
& = \frac{\mp \im g_s^2}{2}\!\! \int \frac{\dfbar^{d-2}q_\perp}{\mb q_\perp^2} \mb \cI_n^{A_j}\! \brac{q_\perp, p_1, p_2} \mb T_j^{A_j} \nn ,
\end{align}
where $\mb \cI_n$ is defined to be the collinear blob with $q^-$ multipole expanded away and $q^+$ integrated over.
Note that, we also set $\bn = n_j$.
Here the upper (lower) sign corresponds to $j \geq 4$ ($j = 3$) and $\Delta_j =\pm q_\perp^2/n\cdot p_j\sim Q\lambda^2$.
The first equality is obtained by avoiding the pole in $q^0$ associated with the leg $j$.
Since $q^+$ is pinched in the upper subgraph, the result for $\mb \cI_n$ is independent of this choice and can be equivalently obtained by simply integrating over $q^+$. Thus, when a soft subgraph is absent, we find a structure of non-collinear leg color factors that exactly mirrors the tree-level situation:
\begin{align}
\spl^{(1)} &= \frac{\im \pi}{\eps} \sum_{j>2} \text{sign}(s_{2j}) \mb T_2 \cdot \mb T_j + \ldots \\
&= \frac{\im \pi}{\eps} \mb T_2 \cdot \big[-2 \mb T_{\rm in} + \underbrace{\brac{ -\mb T_{1} -\mb T_2}}_{-\mb T_{\tilde P}} \big]\, .\nn
\end{align}
Here, we exploited color conservation to remove the sum over outgoing non-collinear legs. This step is important since the second piece corresponding to $-\mb T_{\tilde P}$ is precisely the imaginary part of the Wilson-line emission matrix element with the standard outgoing prescription, and is already included in the collinear matrix elements without Glauber operator insertions.
From \eq{coll_generic}, we can see that the same manipulation hold for arbitrary subgraph.

Note that going to the final equation does require a regularization procedure (either a non-analytic regulator or via taking discontinuities). This is not unexpected since without a soft subgraph, this graph does not have a Glauber region as the $q^-$ (or $q^z$) in the lower part is not pinched and can in fact be deformed into the collinear region $q^- \sim \lambda^2 Q \ra Q$ making $q^+q^- \sim q_\perp^2 \sim Q^2\lambda^2$. This observation is also made recently in \Refcite{Becher:2026kbr} and is explained in terms of being careful with the $\im0$ prescription of the collinear Wilson line differing for incoming and outgoing non-collinear partons.
Crucially, no such regulator is needed in performing $q^+$ integration in $\mb \cI_n$ enabling us to derive the all-order formula in \eq{coll_def}.

As an aside, we now show how adding the non-analytic regulator creates an artificial pinch.
We note that~\cite{Rothstein:2016bsq}
\begin{align}
&\int_{-\infty}^{\infty}\dfbar \ell^z\: \frac{|\ell^z|^{-2\eta}(\nu/2)^{2\eta}}{(2\ell^z + 2A \pm \im 0)} \\
\nn
&= \int_{0}^\infty \frac{\df\ell^z}{4\pi} \: \frac{\brac{2\ell^z/\nu}^{-2\eta}2A}{(\ell^z
+ A \pm \im 0) ( -\ell^z + A \pm \im
0)}\nn \\
&= \int_{-\infty}^\infty\frac{\df\ell^z}{4\pi} \: \frac{A \brac{1 + \cO(\eta)} }{(\ell^z
+ A \pm \im 0) ( -\ell^z + A \pm \im
0)}= \mp\frac{\im}{4} \nn \,.
\end{align}
In going to the third line, we could drop the regulator since the $\ell_z$ is now convergent for $\ell_z\ra \infty$, and extend the integral range to $-\infty \ra \infty$. Thus, we have created an artificial pinch giving the correct Glauber phase term.

\sectionPaper{Cancellation of the zero bins and Lipatov-vertex graphs}
We begin with the computation of the one-loop soft subgraph result in \eq{S1}. Let us consider,
\begin{align}
\includegraphics[height=1.5cm,valign=c]{soft_rescatter_jk.pdf} : \: g_s^2\mb T_2 \cdot \mb T_{\rm in}\mb T_k\int \!\dfbar q^- \dfbar^d \ell (\cI_{k} - \cI_{k}^{(0)}) ,
\end{align}
where the $\cI_k^{(0)}$ is the zero bin of the naive integrand:
\begin{align}
\cI_k = \frac{n_{k\perp} \cdot \brac{q_\perp + \ell_\perp} - (n_k^+/\ell^+ )\ell_\perp\cdot\brac{q_\perp + \ell_\perp}}{(\ell^+/\nu^+)^{-\eta}[\ell^2][-\ell\cdot n_k] [\brac{q+\ell}^2 ] [-q^- -\ell^-]} \, .
\end{align}
Here, all the $[\ldots]$ terms have a `$+\im0$'. For the naive integrand, the $q^-$ pole is pinched which leads to the result in \eq{S1}. This would not happen for correlations between outgoing Wilson lines.
However, we also have zero bins where we set $q^-\!+\!\ell^-\ra0$ or $n_k\cdot \ell \ra 0$ in $\cI_k$. In each case, this leads to a single propagator containing $q^-$ or $n_k\cdot \ell$ resulting in a
Glauber phase with differing signs for incoming and outgoing Wilson lines precisely of the form in \eq{coll_generic}.
Nevertheless, there are also corresponding graphs with the Lipatov vertex that mirror this structure, and they essentially cancel each other. For example, for the $q^- +\ell^- \ra 0$ zero bin, we have
\begin{align}\label{eq:S1zero}
& \includegraphics[height=1.6cm,valign=c]{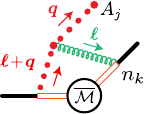}
-\includegraphics[height=1.6cm,valign=c]{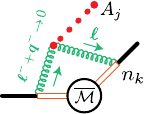}
= -2g_s^2 \mb T_2\cdot \mb T_{\rm in} \mb T_k \times \\
& \nn \int_0^\infty\!\! \frac{\df\ell^+ }{\ell^+}\!\! \int\!\! \frac{\dfbar^{d-2} \ell_\perp}{(\ell^+/\nu^+)^{-\eta} }\!\!\brac{\!
\frac{2r_\perp^2}{\ell_\perp^2 \brac{r_\perp + \ell_\perp}^2}\!+ \!\frac{1}{\brac{q_\perp + \ell_\perp + r_\perp}^2}\!}\!,\nn
\end{align}
where we have defined $r_\perp^\mu \equiv n_{k\perp}^\mu (\ell^+/n_k^+)$.
Thus, we see that while the second term in the last line is scaleless in the transverse integral, the first one is scaleless in rapidity $\ell^+$ integral. The same is also true when $n_k\cdot \ell \ra 0$.

Crucially, we note that this cancellation is independent of the specific leg the bottom Glauber rung is attached to, and since there is no CFV when all the legs are outgoing, we expect this to hold more generally at all-loops. In other words, we expect graphs of type (e) \fig{blobs} to cancel against the zero-bins of the corresponding naive graphs.
Finally, when we consider single Wilson line graphs of type (f) in \fig{blobs}, they lead to a form similar to \eq{S1} but with $n_{k\perp} \ra 0$. Consequently, they result in integrals scaleless in the rapidity.

\end{document}